\documentclass[a4paper,10pt]{article}
\usepackage{jcappub}
\usepackage[T1]{fontenc} 
\usepackage{hyperref}
\input psfig.sty
\pdfoutput=1 
\usepackage{graphics}
\usepackage{graphicx,epsfig,youngtab}
\usepackage{epstopdf}

\def\ba{\begin{eqnarray}}
\def\ea{\end{eqnarray}}

\title{Probing the Cosmological Principle in the counts of radio galaxies at different frequencies}

\author[a]{Carlos A. P. Bengaly Jr.,}
\author[a,b]{Roy Maartens,}
\author[a,c]{Mario G. Santos}

\affiliation[a]{Department of Physics \& Astronomy, University of the Western Cape,\\
Cape Town 7535, South Africa}
\affiliation[b]{Institute of Cosmology \& Gravitation, University of Portsmouth,\\
Portsmouth PO1 3FX, United Kingdom}
\affiliation[c]{SKA South Africa, The Park, 
Pinelands 7405, South Africa}

\emailAdd{carlosap87@gmail.com}

\abstract{
According to the Cosmological Principle, the matter distribution on very large scales should have a kinematic dipole that is aligned with that of the CMB.
We determine the dipole anisotropy in the number counts of two all-sky surveys of radio galaxies. For the first time, this analysis is presented for the TGSS survey, allowing us to check consistency of the radio dipole at low and high frequencies by comparing the results with the well-known NVSS survey. We match the flux thresholds of the catalogues, with flux limits chosen to minimise systematics,  and adopt a strict masking scheme. We find dipole directions that are in good agreement with each other and with the CMB dipole. In order to compare the amplitude of the dipoles with theoretical predictions, we produce sets of lognormal realisations. Our realisations include the theoretical kinematic dipole, galaxy clustering, Poisson noise, simulated redshift distributions which fit the NVSS and TGSS source counts, and errors in flux calibration. The measured dipole for NVSS is $\sim\!2$ times larger than predicted by the mock data.
For TGSS, the dipole is almost $\sim\! 5$ times larger than predicted, even after checking for completeness and taking account of errors in source fluxes and in flux calibration. 
Further work is required to understand the nature of the systematics that are the likely cause of the anomalously large TGSS dipole amplitude.
}

\date{\today}


\begin{document}
\maketitle
\flushbottom

\section{Introduction}

A fundamental hypothesis of the current standard model of cosmology is statistical homogeneity and isotropy, i.e., the absence of privileged positions and directions in the Universe on sufficiently large scales. This is the Cosmological Principle and it applies for fundamental observers, who see no kinematic dipole in the cosmic microwave background (CMB) radiation. Our velocity relative to fundamental observers is deduced from the dipole anisotropy in the CMB temperature, which corresponds to~\citep{kogut93, planck14}
\ba\label{beta}
\beta \equiv {v \over c} = 1.23\times 10^{-3} \quad \mbox{towards the direction}~~ (l,b) = (264^{\circ},48^{\circ})~~\mbox{in galactic coordinates.}
\ea

In the standard cosmological model, this dipole arises from the Doppler boost  of CMB photons due to our relative peculiar velocity\footnote{There are alternative interpretations of the CMB dipole  as an intrinsic property arising from, e.g.,  primordial density perturbations~\cite{roldan16}, or our location near the centre of a void~\citep{cusin16}.}, so that we should observe statistical isotropy once this signal is taken into account. Analyses have been performed using CMB observations and are consistent with isotropy.\footnote{Some potential departures from isotropy have been reported, but the statistical significance and possible physical explanation remain open issues~\citep{schwarz16, planck16a}.} 

A critical requirement of statistical isotropy is that the Solar System rest frame seen in the CMB and in the number counts of distant radio sources should be consistent. A test of this was proposed in~\cite{ellis84}, which estimated the dipole amplitude we should see in the projected source distribution, given the velocity inferred from the CMB dipole, and taking into account  the Doppler-boosted frequency of sources and the aberration of their locations in the sky. The first implementation of this test was carried out in~\cite{baleisis98}, which combined the Green Bank 1987 (87GB) and Parkes-MIT-NRAO (PMN) catalogues, but due to the limited number of sources ($\sim 40,000$), it was not possible to obtain a statistically significant signature of our velocity. This was obtained by~\cite{blake02}, which found a $>2\sigma$ detection of our relative motion in the NRAO VLA Sky Survey (NVSS), with roughly 5 times more sources. The result was consistent with the CMB dipole within $2\sigma$ confidence level (CL) for both amplitude and direction. 

The analysis was reassessed in~\cite{singal11}, which obtained a similar direction to~\cite{blake02}, but the amplitude was consistent with a peculiar velocity of $v \sim 1500 \, \mathrm{km/s}$, i.e., much larger than that seen in the CMB. Moreover,~\cite{singal11} found that this result could not be reproduced at random with $>99\%$ CL. Other analyses of the NVSS data were performed~\citep{gibelyou12, rubart13, tiwari14,fernandezcobos14, tiwari15}, with most of them confirming the results of~\cite{singal11}. The exceptions were~\cite{tiwari16a, colin17}, which pointed out that the large radio dipole could be partially ascribed to the local structure, thus reducing the tension between the observed and the expected dipole signal from $\sim 3\sigma$ to $(2.2-2.8)\sigma$, depending on the method and the flux cut-off adopted. Some alternative explanations for this result, such as the existence of a large void in the local Universe~\citep{rubart14}, or super-horizon anisotropic modes from primordial perturbations~\citep{ghosh14}, were tested against observations. However, none of them could reconcile the expected and observed dipole amplitudes unless strong fine-tuning was applied. Other analyses applied to the NVSS catalogue, such as the two-point angular correlation function of its radio sources~\citep{chen16}, cross-correlations between NVSS and the WMAP CMB experiment~\citep{monteagudo10}, and tests of the isotropy of the slope of the cumulative number source counts~\citep{ghosh17}, on the other hand, revealed no significant departure from the standard cosmological model assumptions. 

Here we perform a new assessment of the cosmic radio dipole. Using the new all-sky continuum survey TGSS (TIFR GMRT Sky Survey), we probe for the first time the dipole in a different all-sky survey than NVSS. More importantly, we perform a comparison of the dipoles found at different frequencies, given that TGSS operates at a much lower frequency band than NVSS. We also follow the idea of~\cite{tiwari16a} and test the hypothesis that a large dipole is related to local structure, by simulating sky maps for both surveys, using a simulated redshift distribution and assuming a galaxy bias. 

The paper is organised as follows: section 2 describes the observational data and its processing; section 3 discusses the methodology adopted; section 4 presents the results and, finally, our discussion and concluding remarks are given in section 5.

\section{The observational data}

\begin{figure*}[!ht]
\includegraphics[width = 5.5cm, height = 8.0cm, angle=90]{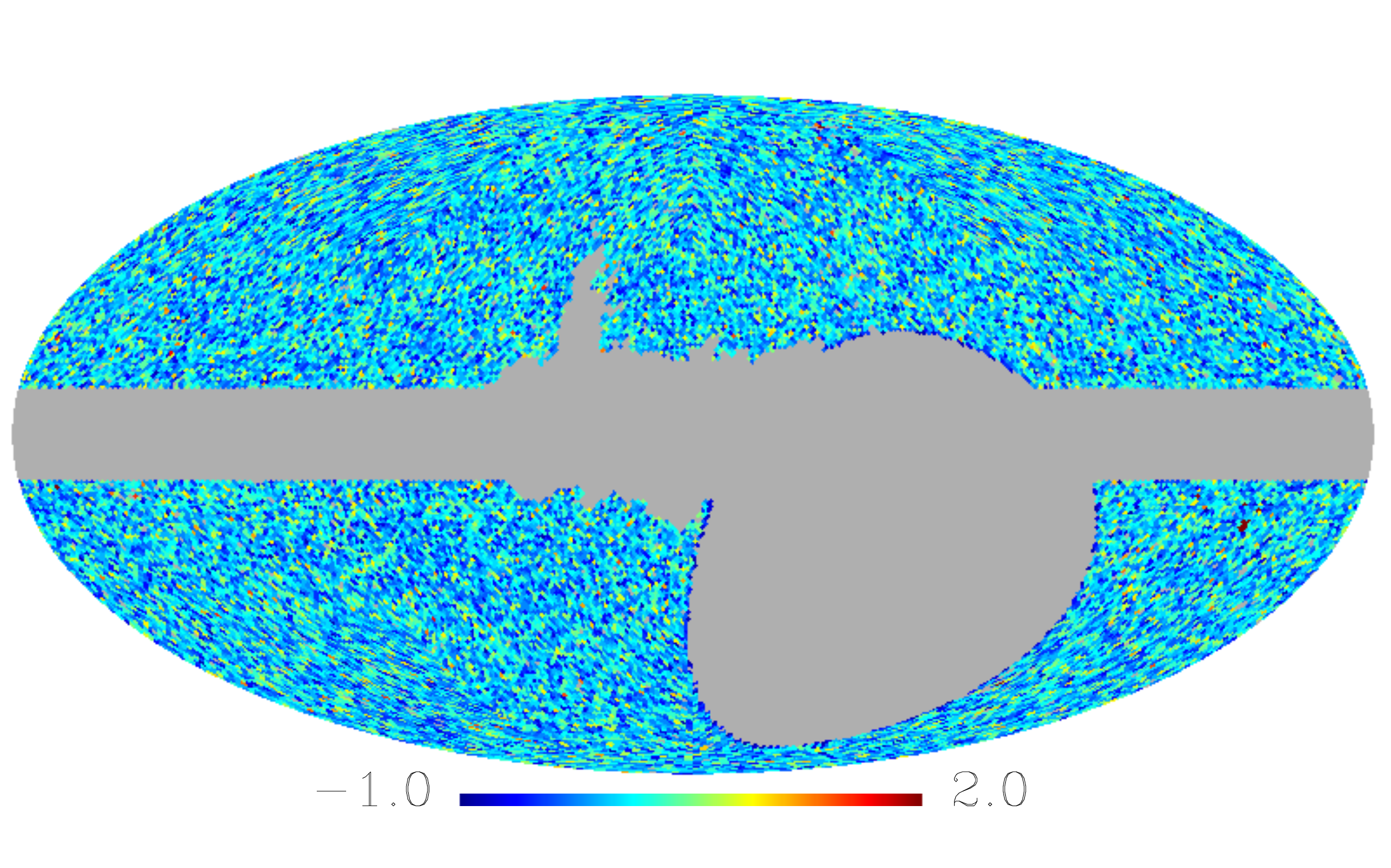}
\includegraphics[width = 5.5cm, height = 8.0cm, angle=90]{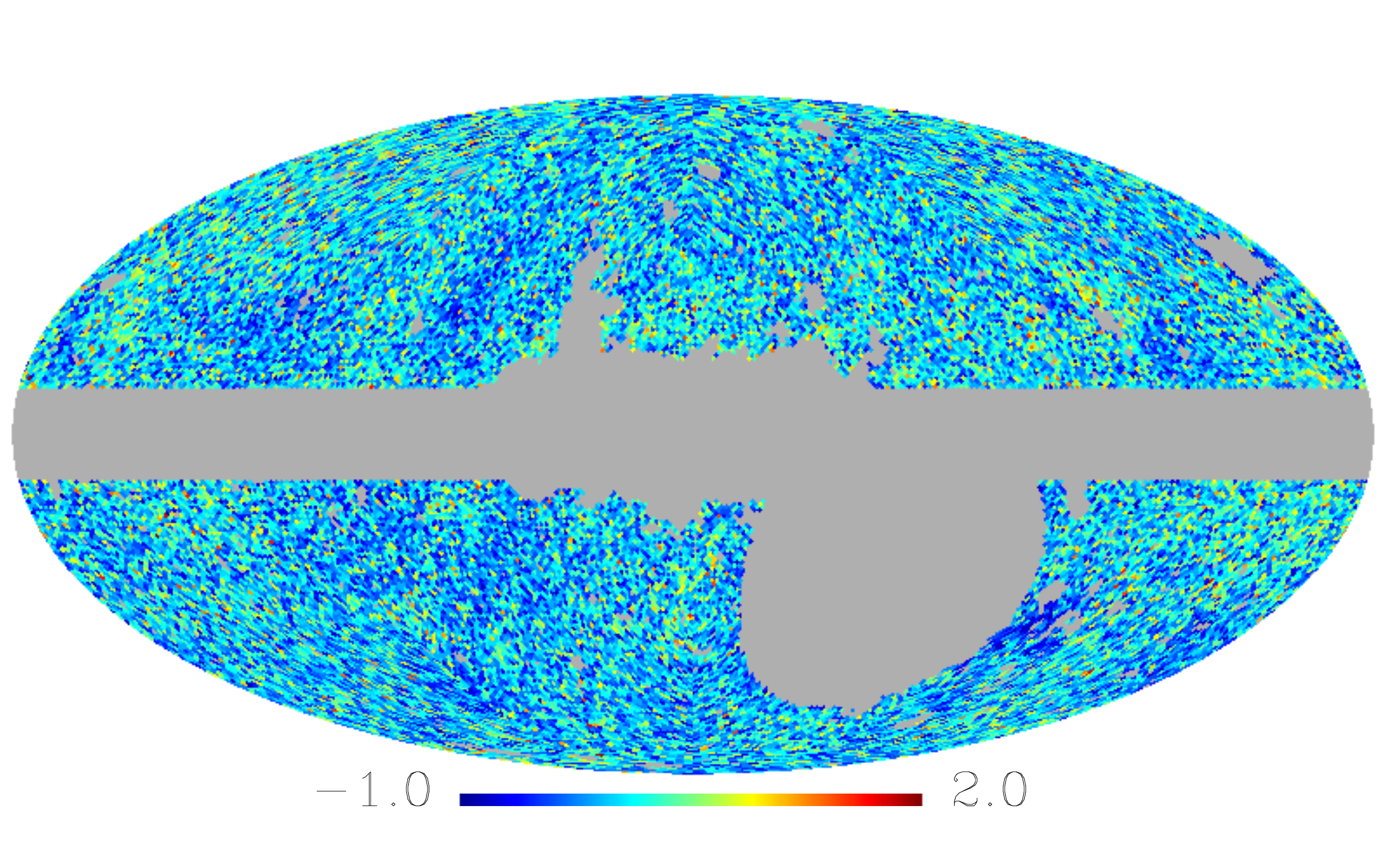}
\caption{Pixelised number count map of NVSS (left) and TGSS (right) radio sources in the flux ranges $20<S<1000 \, \mathrm{mJy}$ and $100<S<5000 \, \mathrm{mJy}$, respectively. Number counts are normalised to the average count per pixel, and clipped at $2.0$ to ease visualisation.} 
\label{fig:nummap_realdata}
\end{figure*}

The radio continuum data  consists of two surveys: NVSS, conducted by the VLA at high frequency, 1.4\,GHz~\cite{condon98}, and TGSS, conducted by GMRT at low frequency, 150\,MHz~\cite{intema17}. Both cover a large area of the sky ($\mathrm{DEC}  > -40^{\circ}$ in NVSS and $\mathrm{DEC} > -53^{\circ}$ in TGSS), so they are suitable to carry out isotropy tests. In order to compare the number count dipoles, we match their flux ranges using $S \propto \nu^{\alpha}$, where $\alpha=0.76$~\citep{tiwari16b} is the spectral index, so that
\begin{equation}
\label{eq:S_alpha}
S_{\rm NVSS} = S_{\rm TGSS} \left({\nu_{\rm TGSS}\over \nu_{\rm NVSS}}\right)^{\alpha} \simeq 0.20 \, S_{\rm TGSS} \,.
\end{equation}
\underline{Flux selection}:
\begin{itemize}
\item
We choose $S_{\rm TGSS}>100 \, \mathrm{mJy}$ since  TGSS is complete in this range, as shown in Fig. 20 of~\cite{intema17}. 
\item
The average rms noise across the map is $\simeq$ 4.0 mJy/beam. However, this noise can have strong fluctuations (up to 2000  mJy/beam). In order to have a more uniform noise across the image, we mask pixels with rms noise above 10 mJy/beam, which will remove the outliers. This will match well with our imposed flux cut of 100 mJy/beam (e.g. at least $10\sigma$ above the rms noise).
\item
In order to remove the brightest sources we take  $S_{\rm NVSS}<1000 \, \mathrm{mJy}$. 
\item
$100<S_{\rm TGSS}<5000 \,$mJy  roughly corresponds to $\, 20<S_{\rm NVSS}<1000 \,$mJy. 
\end{itemize}
\underline{Mask:}\\
In order to purify the TGSS and NVSS catalogues, we adopt a rigorous masking procedure to deal with well-known contamination. This involves the elimination of pixels in the following regions:

\begin{itemize}

\item Close to the galactic plane, i.e., $|b| \leq 10^{\circ}$.

\item Within $1^\circ$ of the local radio sources given in~\cite{vanvelzen12}.

\item Within $1^\circ$ of local superclusters given in Table 1 of~\cite{colin17}.

\item Galactic foreground emission above $T=50\, \mathrm{K}$ according to the 408 MHz continuum map in~\cite{haslam82}.
 
\end{itemize}
The final TGSS and NVSS maps are shown in Fig.~\ref{fig:nummap_realdata} for $100<S_{\rm TGSS}<5000 \, \mathrm{mJy}$ and $20<S_{\rm NVSS}<1000 \, \mathrm{mJy}$, at a resolution of $N_{\rm side}=64$, so that each pixel is $\sim 55'$. The remaining sky area, numbers per pixel and total numbers after the masking procedure just described are:
\ba
\mbox{TGSS}\quad && f_{\rm sky} \simeq 0.687, ~~\bar{n} \simeq 6.912, ~~N_{\rm total} = 233,395\\
\mbox{NVSS}\quad && f_{\rm sky} \simeq 0.657, ~~\bar{n} \simeq 7.848, ~~N_{\rm total} = 253,313      
\ea

\section{Estimator and uncertainties}

\subsection{Dipole estimator}

The radio dipole anisotropy is estimated by decomposing the sky into 3072 HEALPix cells ($N_{\rm side}=16$)~\cite{gorski05}, and counting the difference in the number of sources contained in opposite hemispheres whose symmetry axes are provided by these cell centres. Therefore, we can construct an estimator, hereafter called the `delta-map', defined by 
\begin{equation}
\label{eq:delta}
\Delta(\theta) \equiv \frac{\sigma_i^U(\theta) - \sigma_i^D(\theta)}{\sigma} = A_{\rm obs} \cos\theta \quad \mbox{where}\quad \sigma_i^J = {N^J_i \over 2\pi (f_{\rm sky})_i^J}\,, ~~\sigma = {N_{\rm total}\over 4\pi f_{\rm sky}}\,.
\end{equation}
Here $i=1,\cdots, 3072$ labels the hemisphere decomposition, and $J=U,D$ identifies the `up' and `down' hemispheres in this pixelisation scheme; $N^J_i$, $(f_{\rm sky})^J_i$, and $\sigma_i^J$  are the number of sources, the observed sky fraction, and the source density in each hemisphere, respectively, with $N^U_i + N^D_i = N_{\rm total}$. The angle $\theta$ corresponds to the position of the $i$-th pixel centre in the sky with respect to the estimated direction of our relative motion, and thus the left-hand side of \eqref{eq:delta} provides the dipole amplitude of the source count, $A_{\rm obs}$, in this direction. 

\begin{figure*}[!ht]
\includegraphics[scale=0.36]{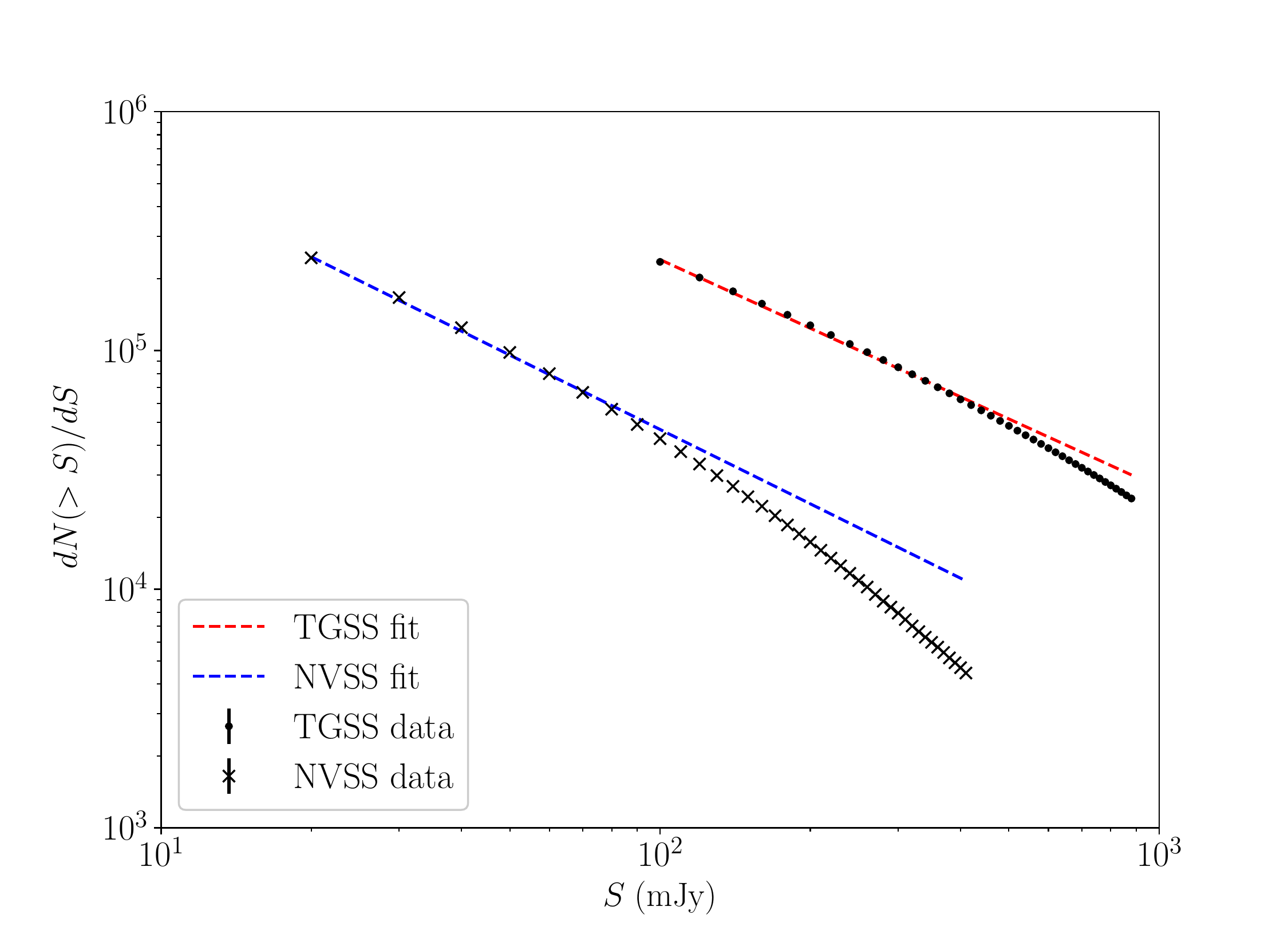}
\includegraphics[scale=0.39]{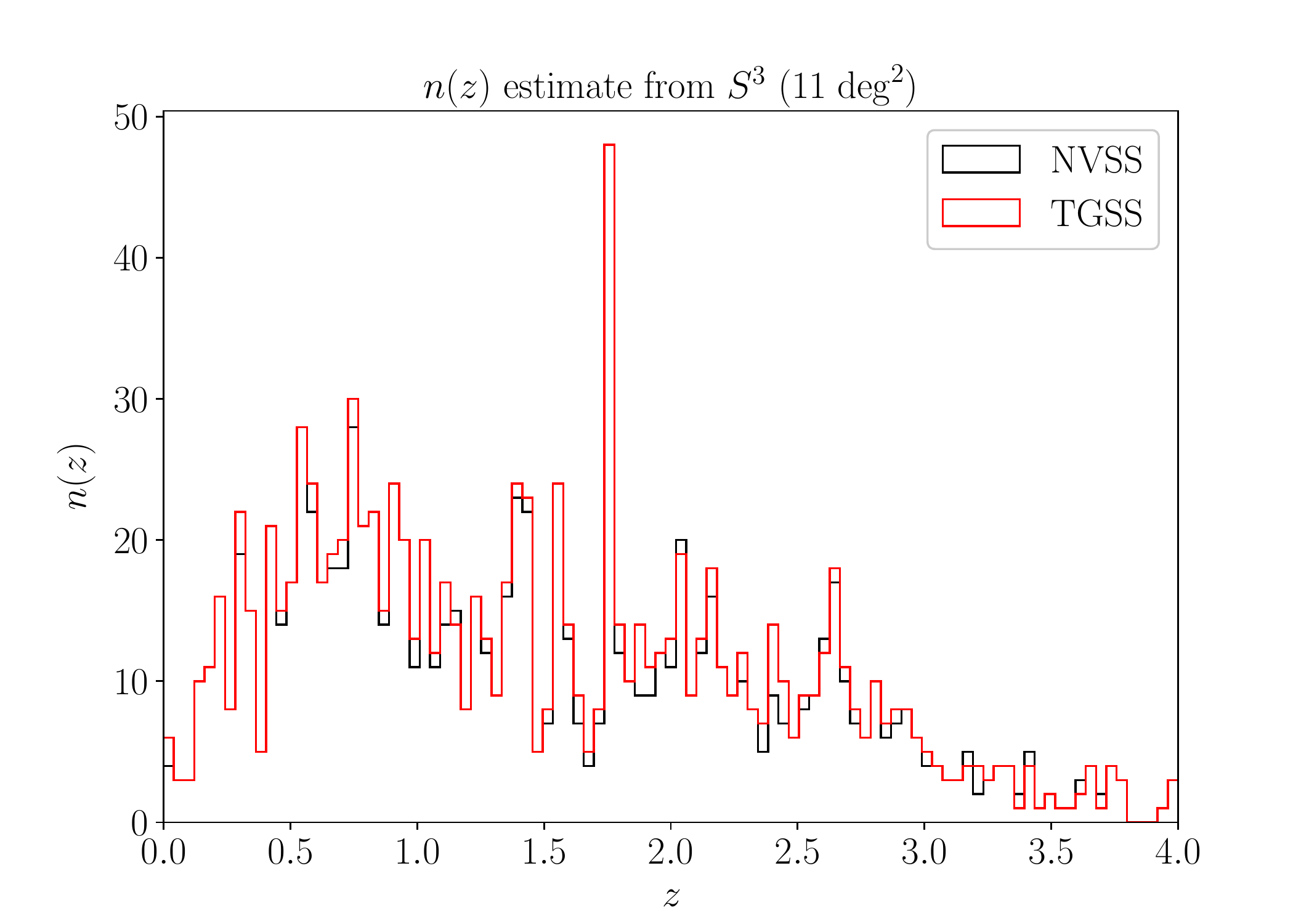}
\caption{Source counts \eqref{eq:dN_dS} per flux bin per solid angle for TGSS and NVSS data, compared to their best fits (left) (the error bars are too small to be visible); the redshift distribution $n(z)$ for both catalogues, based on $S^3$ simulations (right).}
\label{fig:logn_logs_nz}
\end{figure*}

We can compare this quantity with the expected signal due to Doppler and aberration effects~\cite{ellis84,gibelyou12,rubart13}:
\begin{equation}
\label{eq:A_kin}
A_{\rm kin} = \big[2 + x(1+\alpha)\big]\beta \,,
\end{equation}
where $x$ is the power law index, defined by
\begin{equation}
\label{eq:dN_dS}
N(>S) \propto S^{-x} \,.
\end{equation}
The plots in the left panel of Fig.~\ref{fig:logn_logs_nz} show that 
\ba
x \simeq 1.015 \pm 0.011 ~~ \mbox{for NVSS}, \qquad x \simeq 0.955 \pm 0.010 ~~ \mbox{for TGSS.} 
\ea
Together with \eqref{beta}, this leads to 
\begin{equation}
\label{eq:dip_amp_nvss}
A_{\rm kin} \simeq 0.00466 \quad\mbox{towards the direction }~ (l,b) = (264^{\circ},48^{\circ}) \quad\mbox{for NVSS},
\end{equation}
\begin{equation}
\label{eq:dip_amp_tgss}
A_{\rm kin} \simeq 0.00453 \quad\mbox{towards the direction }~ (l,b) = (264^{\circ},48^{\circ}) \quad\mbox{for TGSS},
\end{equation}
This will be hereafter regarded as the expected radio dipole signal if there is consistency with the Cosmological Principle.

\subsection{Mock data}

The statistical significance of the observed dipole $A_{\rm obs}$ is estimated through 1,000 lognormal realisations of the galaxy density field produced by the {\sc flask} code\footnote{\url{http://www.astro.iag.usp.br/~flask/}}~\cite{xavier16}. These mock data-sets assume the angular power spectrum $C_\ell$ as provided by {\sc camb sources}\footnote{\url{http://camb.info/sources/}}~\cite{challinor11}, using the latest {\em Planck} best-fit $\Lambda$CDM as the fiducial cosmology~\cite{planck16b}. The power spectrum is computed in five redshift bins over $0 < z < 4$, with a top-hat window and bin edges at $z = 0.0, 0.5, 1.0, 2.0, 3.0, 4.0$. We use the redshift distribution $n(z)$ shown in the right panel of Fig.~\ref{fig:logn_logs_nz}. 

In order to obtain $n(z)$, we ran a query on the S$^3$ (SKA Simulated Skies) website\footnote{\url{http://s-cubed.physics.ox.ac.uk/s3_sex}} (see~\cite{wilman08} for details of the simulation). By assuming a sky area of $11 \,\mathrm{deg}^2$, the frequency of both surveys, as well as the specified flux range for each frequency, we obtained ~1,000 matches for them whose redshift distribution of sources is exhibited in the right panel of Fig.~\ref{fig:logn_logs_nz}. These values correspond to an average source density of $\sim 10 \,\mathrm{sources}/\mathrm{deg}^2$, so that we normalise the source density of the mocks according to the $n(z)$ to match both $\bar{n}$ and $N_{\rm total}$ of the real data once we apply their masks, and Poisson-sample the number count fields across the available sky area. 
For the galaxy bias, we adopt the model given by~\cite{tiwari16a}:  
\ba
b(z) = 1.6 + 0.7z + 0.35z^2 \,,
\ea 
but we relaxed their assumption that $b(z>1.5)=b(z=1.5)$, in line with the argument given in~\cite{Ferramacho:2014pua}.

In order to simulate the kinematic dipole in the mock catalogues, we apply a dipolar modulation  according to~\eqref{eq:dip_amp_nvss} and~\eqref{eq:dip_amp_tgss}.

\subsection{Flux calibration uncertainties}

\begin{figure*}[!t]
\includegraphics[width = 5.5cm, height = 8.0cm, angle=90]{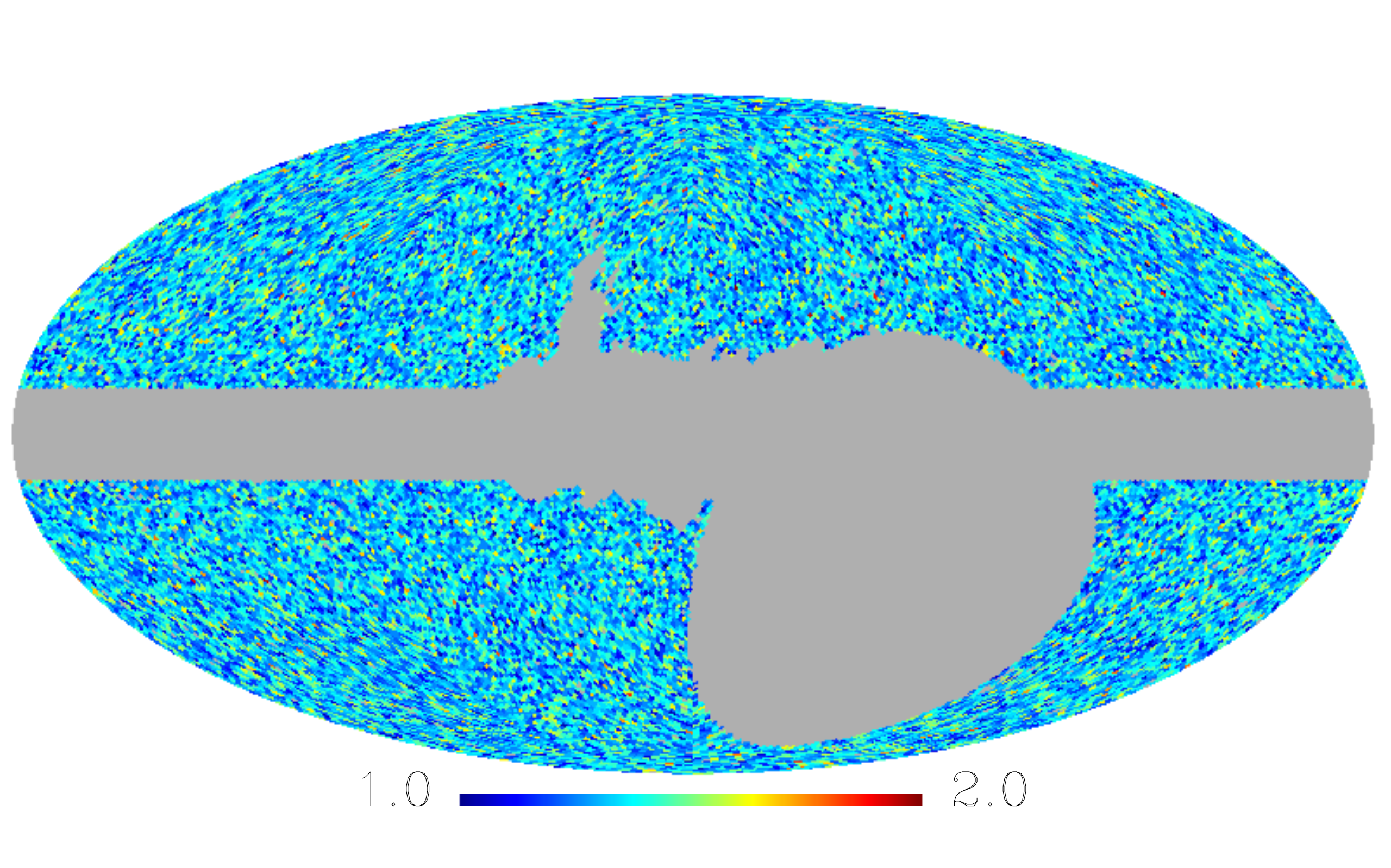}~~~
\includegraphics[width = 5.5cm, height = 8.0cm, angle=90]{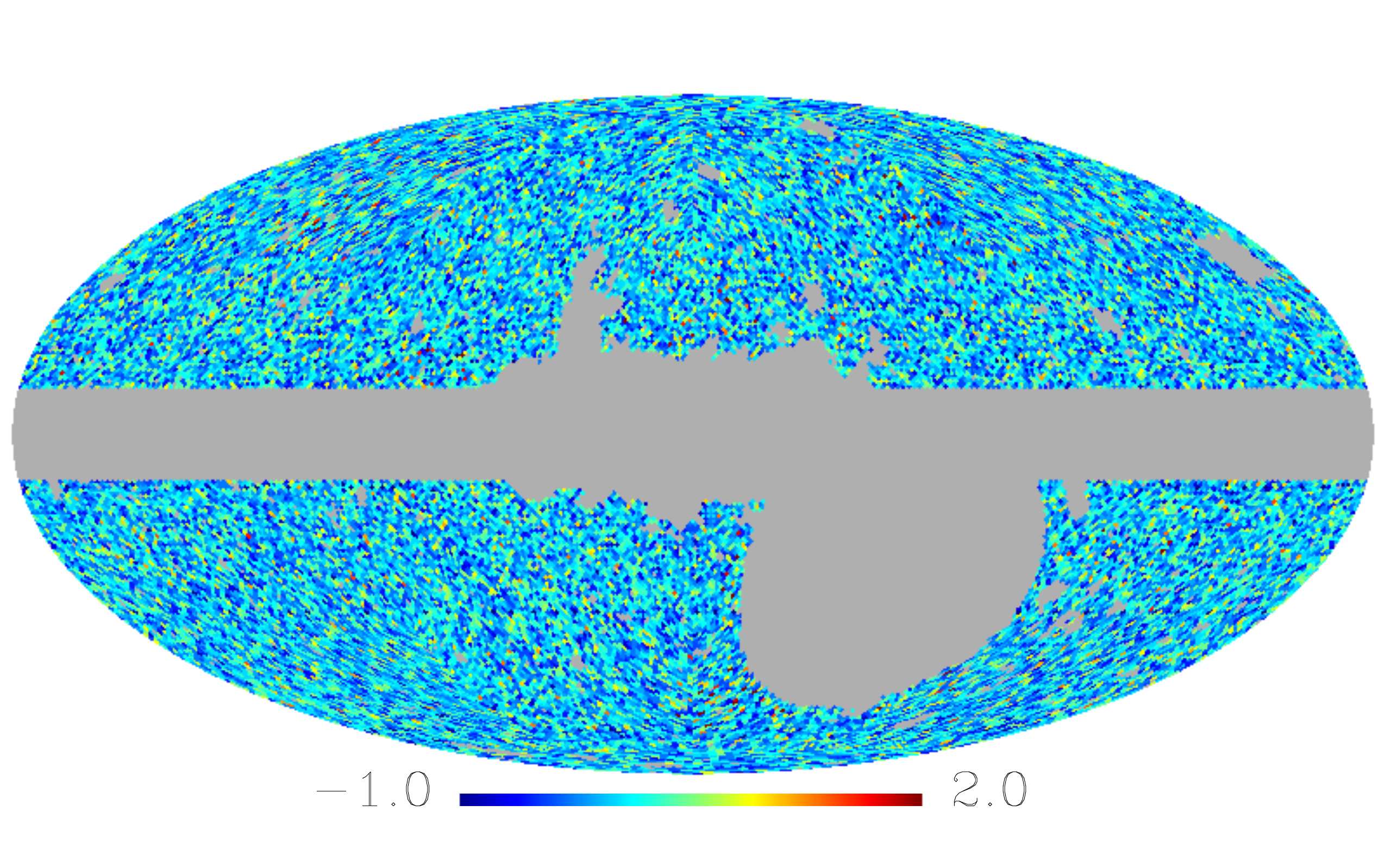}
\caption{NVSS (left) and TGSS (right) mock count maps in the same flux range as Fig.~\ref{fig:nummap_realdata}, including flux calibration corrections.} 
\label{fig:nummap_mockdata}
\end{figure*}

As discussed in~\cite{ska15}, errors in the flux calibration can introduce spurious anisotropies in the number count maps, and hence affect their dipole estimates. For the NVSS catalogue, we model this effect via~\cite{ska15}
\begin{equation}
\label{eq:corr_S}
S_{\rm corr}(\epsilon,\delta) = {S_0\over \big[1 + \epsilon r/\cos{(\delta - \delta_*)}\big]} ,
\end{equation}
where $\epsilon$ determines the amplitude of the flux calibration errors, $0\leq r\leq 1$ is a random variable, $\delta$ is the declination of the sources, $\delta_* = -30^{\circ}$ and $S_0$ is the observed flux.
Then the number counts of the simulations are changed  to
\begin{equation}
\label{eq:corr_N}
N_{\rm corr}(\epsilon,\delta) = N_0\big[1 + \epsilon r/\cos{(\delta - \delta_*)}\big]^x .
\end{equation}
For NVSS, we assume that 
\ba \label{fcn}
\mbox{NVSS:}\qquad\epsilon=0.05\,.
\ea
Low-frequency radio observations face greater difficulties with flux calibration. Recent work~\cite{hurley17} indicates that TGSS has flux calibration errors of $\sim 10\%-20\%$ on few-degree scales. 
TGSS was observed in pointings that are typically clustered together. Flux scale deviations are strongly linked to observing sessions, so that the scales on which deviations are seen relate to the area covered in these pointings, covering $\sim3^{\circ}-30^{\circ}$.\footnote{H. Intema, private communication.} 
We assume that the directional dependence of the flux calibration errors is provided by regions within $10^{\circ}$ of the pointings, instead of the factor $\cos{(\delta - \delta^*)}$ in \eqref{eq:corr_S}. For TGSS, we assume
\ba\label{fct}
\mbox{TGSS:}\qquad\epsilon=0.15.
\ea

As a result of this procedure, we obtain mock NVSS and TGSS catalogues that reproduce the features of the real data, i.e., the shot noise,  the galaxy clustering according to the $\Lambda$CDM matter density perturbations, the $n(z)$ in the considered flux range, the incomplete sky coverage, and the flux calibration errors. Examples of catalogues produced according to this prescription are shown in Fig.~\ref{fig:nummap_mockdata}. 

\subsection{Source flux uncertainties}

The real data comprises sources with large flux uncertainties, i.e. 10\% or larger of their respective values. These flux errors are indicated in the source catalogue and include several uncertainties in the flux extraction process from the image. Moreover, the rms noise is not homogeneous across the image. These effects mean that, when imposing a constant flux cut throughout the image, some of the galaxies might be above or below such a flux cut, simply due to errors in the flux. This effect should be small since we are using a flux cut well above the rms noise. At the same time, this effect should not have any angular scale dependence, so it is not expected to affect the dipole. Nevertheless, we include such an effect in the simulations. We implement this
by drawing the flux value from a Gaussian distribution 
$\mathcal{N}(S,\sigma)$, with  original  flux value $S$ and total flux uncertainty $\sigma $.
The standard deviation is obtained from
\ba\label{eq:sigmaS}
\mbox{TGSS:} \quad  \sigma^2 = \sigma_S^2 + \mbox{(rms noise)}^2\,,\qquad
\mbox{NVSS:} \quad  \sigma = \sigma_S \;,
\ea
where $\sigma_S$ is the flux error. With this process, we can quantify whether the source counts could exhibit additional anisotropies because of the flux errors.

\subsection{Angular power spectra}

\begin{figure*}[!t]
\includegraphics[scale=0.39]{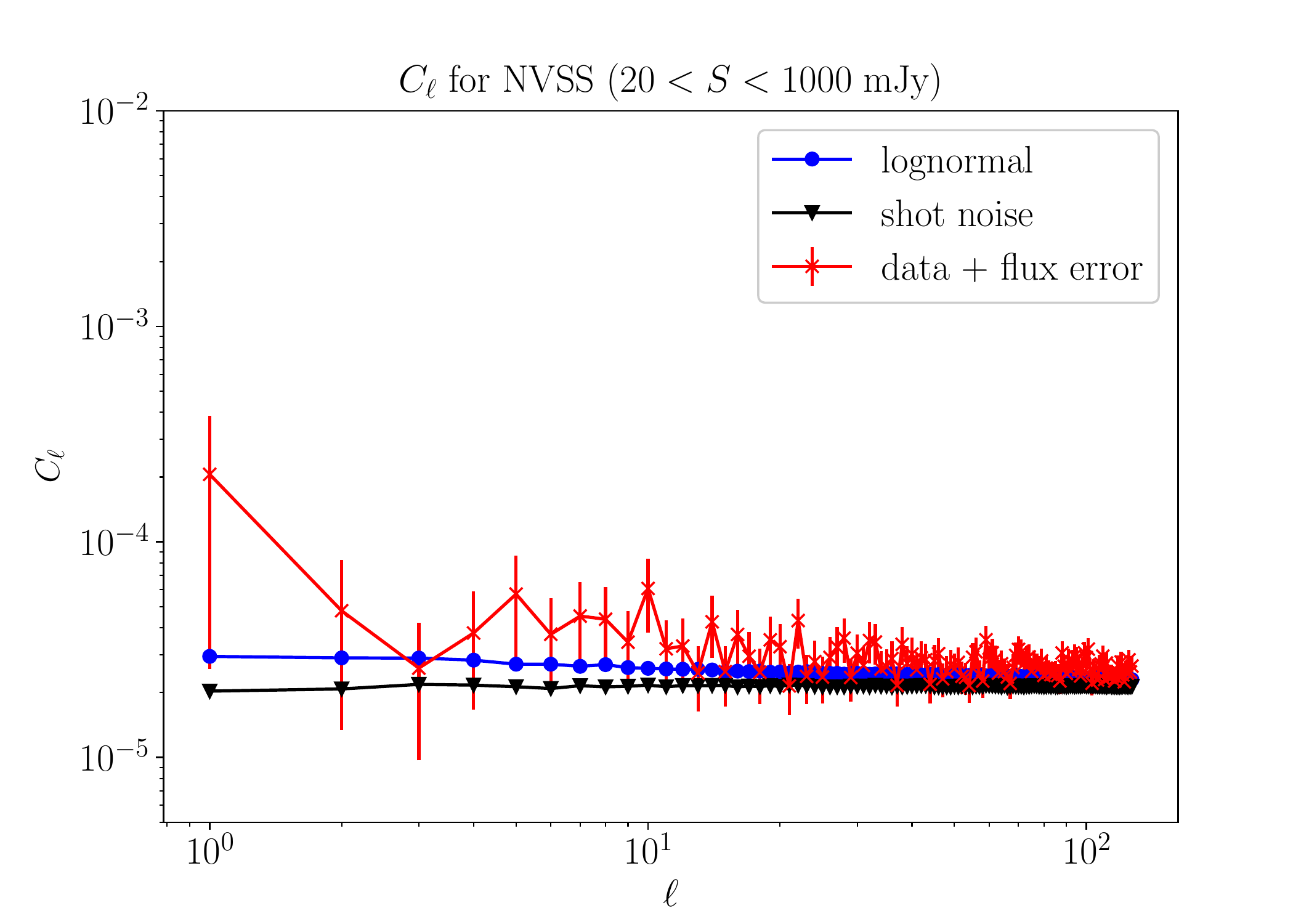}~~~~~~
\includegraphics[scale=0.39]{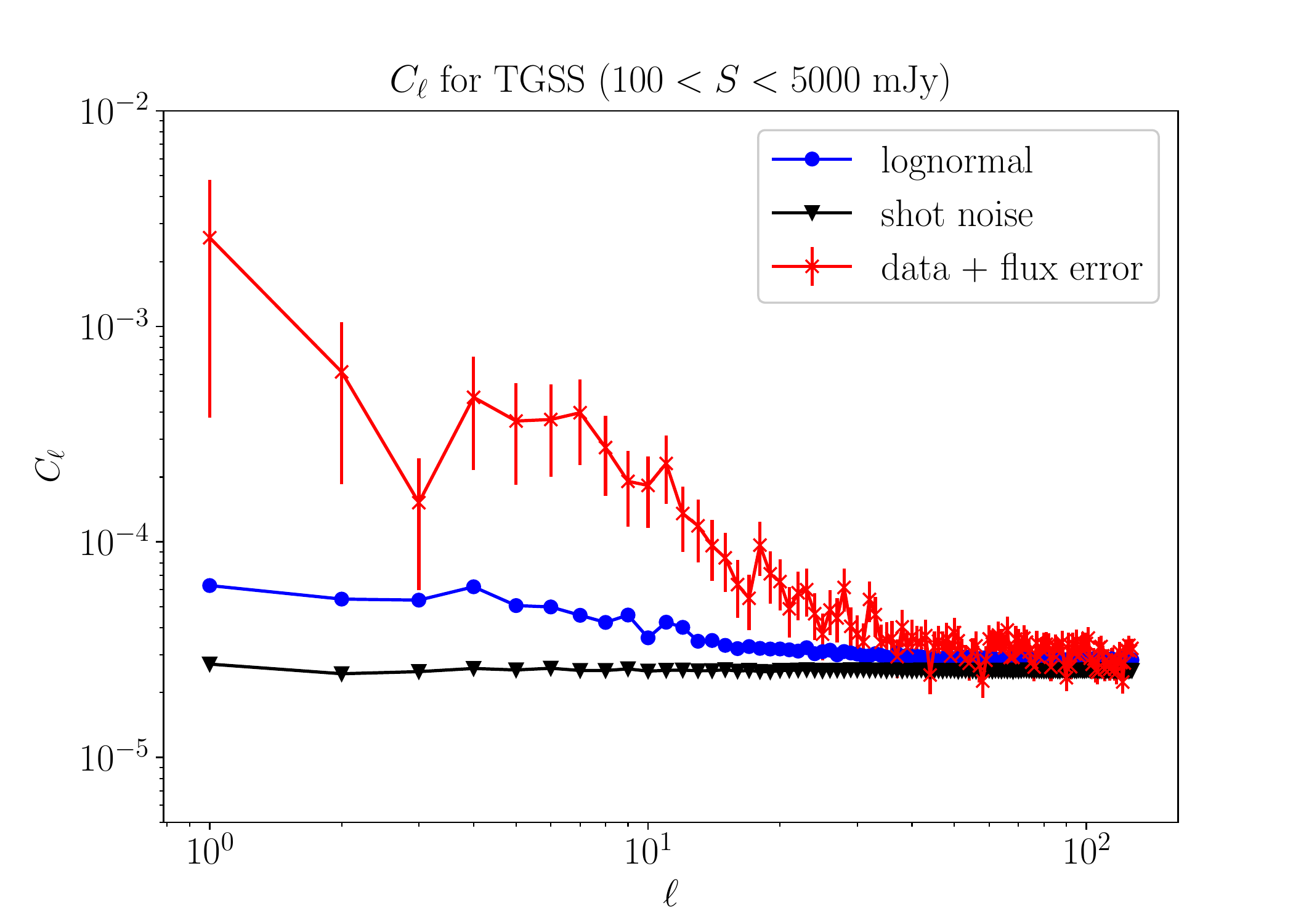}
\caption{Angular power spectra (red) for the NVSS (left) and TGSS (right) maps featured in Fig.~\ref{fig:nummap_realdata}, with uncertainties given by \eqref{cv}. Average spectra for 1,000 NVSS and TGSS mock catalogues (blue), including flux calibration corrections, also shown. Black dots give the shot noise contribution.} 
\label{fig:Cls}
\end{figure*}

Figure~\ref{fig:Cls} shows the angular power spectra of the real maps and of the mock data averages. The uncertainties for the real maps are given by cosmic variance and the standard deviation of the average $C_\ell$ when accounting for the flux uncertainties:
\ba\label{cv}
\sigma_{C_\ell}^2 =\sigma_{\rm cv}^2+\sigma_{\rm flux}^2, \quad \sigma_{\rm cv}^2= {2\over (2\ell + 1)f_{\rm sky}}C_\ell \,.
\ea
The average $C_\ell$ of the mock data-sets including the flux calibration error simulations are also shown, along with the shot noise. We imposed a maximum multipole $\ell_{\rm max} = 128$. 

It is evident that the power spectra are indistinguishable from shot noise on small angular scales for both surveys. The TGSS data shows much larger number density fluctuations than the NVSS, and than those of the mock data, even when cosmic variance and errors in source fluxes and flux calibration are taken into account.

\subsection{Signal to noise estimates}

\begin{table*}[!ht]
\centering
\begin{tabular}{cccc}
\hline
\hline
&&&\\
Flux range (mJy) & $A_{\rm PN}$ & $\mathrm{SNR}$ & $\mathrm{DB}$ \\ &&&\\
\hline
\hline 
$100<S_{\rm TGSS}<5000$ & $0.00427$ & $0.9340$ & $2.4503$ \\ 
\hline
$~20<S_{\rm NVSS}<1000$ & $0.00386$ & $1.0371$ & $2.3222$ \\ 
\hline
\hline 
\end{tabular}
\caption{Flux range, Poisson noise dipole contribution \eqref{apn},  SNR \eqref{eq:SNR} and dipole bias \eqref{eq:DB}. The kinematic dipole amplitude is assumed to be given by \eqref{eq:dip_amp_nvss} and \eqref{eq:dip_amp_tgss}.}
\label{tab:SNR}
\end{table*}

Before proceeding with the delta-map analysis, we can perform a rough assessment of the signal-to-noise ratio (SNR) and the dipole bias (DB) of the kinematic dipole detection, in a similar fashion to~\cite{itoh10} (see also~\cite{baleisis98}):
\ba
\label{eq:SNR}
\mathrm{SNR} &=& \frac{A_{\rm kin}}{\sqrt{A_{\rm LSS}^2 + A_{\rm PN}^2}} \,, \\ 
&&\nonumber \\
\label{eq:DB}
\mathrm{DB} &=& \frac{A_{\rm kin} + A_{\rm LSS} + A_{\rm PN}}{A_{\rm kin}} \,. 
\ea
Here $A_{\rm kin}$ is given by \eqref{eq:dip_amp_nvss} and~\eqref{eq:dip_amp_tgss}, while $A_{\rm LSS}$ is obtained from the average $C_1$ of the 1,000 {\sc flask} realisations, but excluding 
Poisson noise and the dipole modulation $A_{\rm kin}$, so that the only expected contribution to this average $C_1$ is from galaxy clustering. We find that
\ba
A_{\rm LSS} = {3\over2}\sqrt{C_1 \over \pi} \; \simeq 0.00230\,.
\ea
The Poisson noise contribution to the dipole is given by 
\ba\label{apn}
A_{\rm PN} = {3\over2}\sqrt{f_{\rm sky} \over \pi \bar{N}}\,, 
\ea
where $\bar{N}$ is the average number of sources per steradian.

The results of the SNR and DB estimates are displayed in Table~\ref{tab:SNR}. We note that the Poisson noise contribution $A_{\rm PN}$ is comparable to the amplitude of the kinematic dipole $A_{\rm kin}$, for all cases analysed. The large-scale structure contribution $A_{\rm LSS} $ is roughly half of the Poisson noise, and so we obtain $\mathrm{SNR} \simeq 1.0$ and $\mathrm{DB} > 2.0$ for all cases. This means that a kinematic dipole is barely detectable, given the properties of present radio catalogues, and also that the dipole we expect to obtain from the data should be $2.3-2.5$ times larger than the purely kinematic signal. Because of this, we only consider the largest flux ranges available, in order to minimise the effects from the large shot noise $A_{\rm PN}$.

\section{Measured dipoles: real data}

\begin{figure*}[!ht]
\centering
\includegraphics[width = 5.5cm, height = 8.0cm, angle=90]{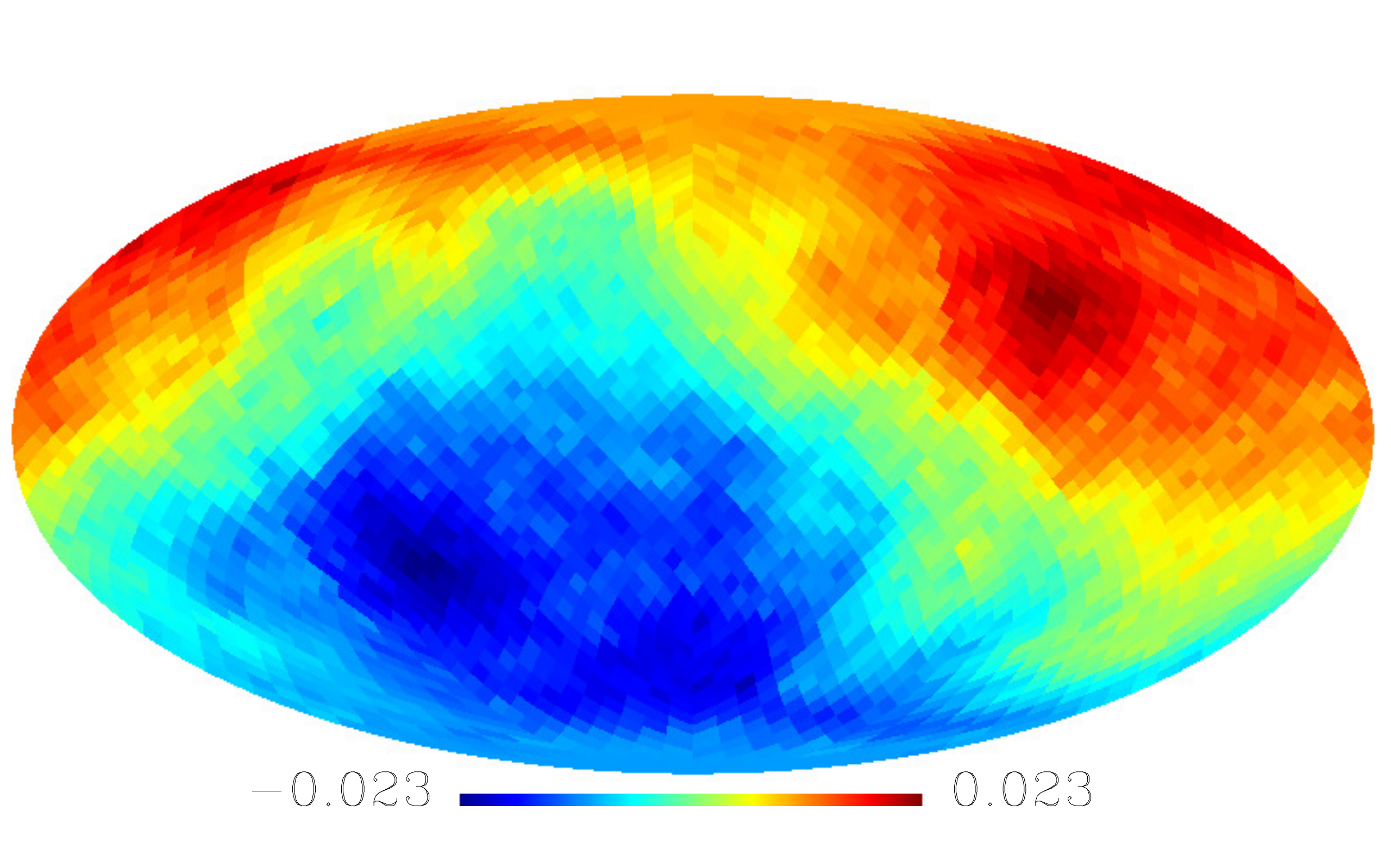}~~
\includegraphics[width = 5.5cm, height = 8.0cm, angle=90]{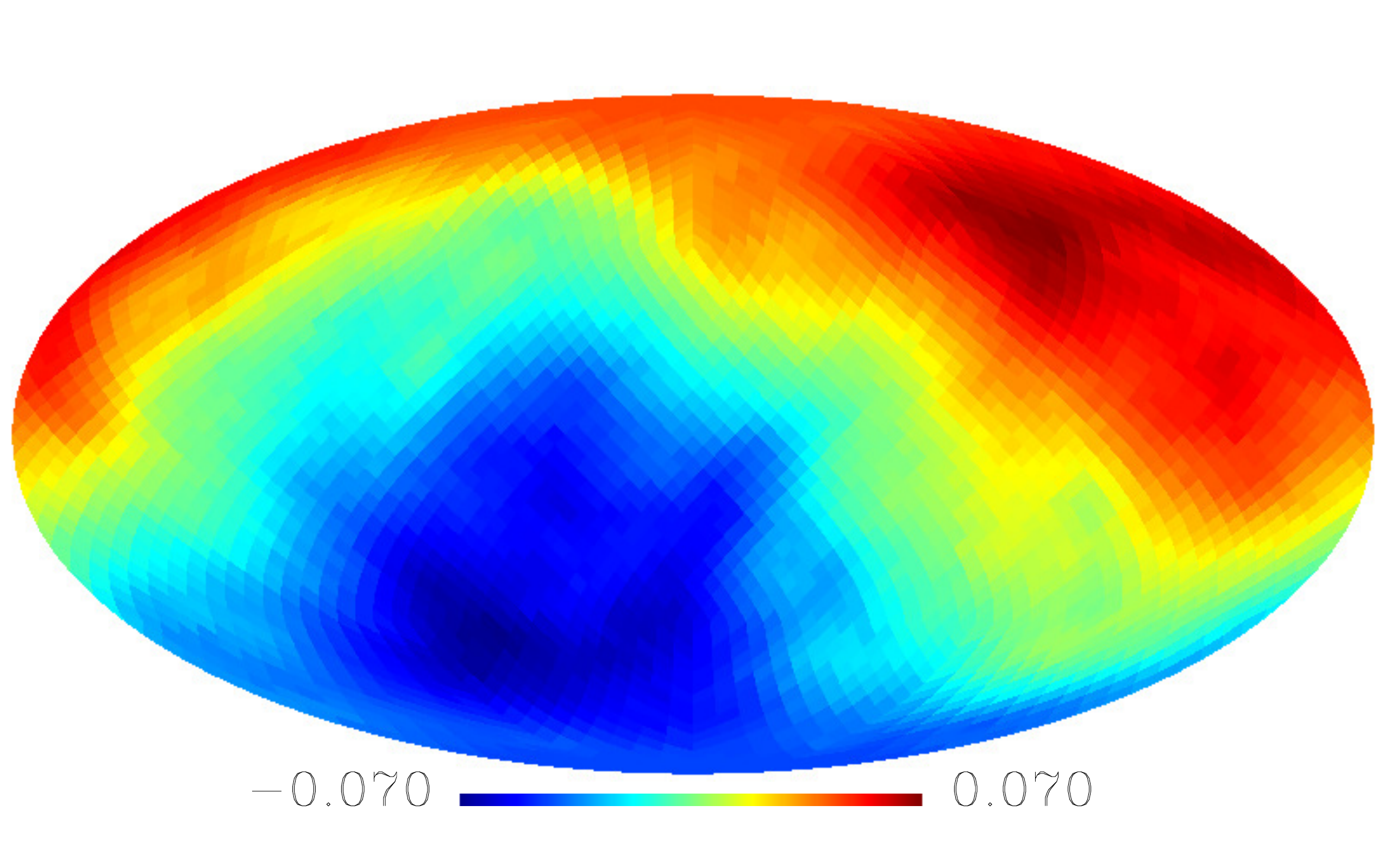}
\caption{Dipole anisotropy of NVSS (left) and TGSS (right) source counts from applying the delta-map \eqref{eq:delta} to the data shown in Fig.~\ref{fig:nummap_realdata}.} 
\label{fig:deltamap_realdata}
\end{figure*}

\begin{figure*}[!t]
\includegraphics[width = 8.0cm, height = 5.5cm]{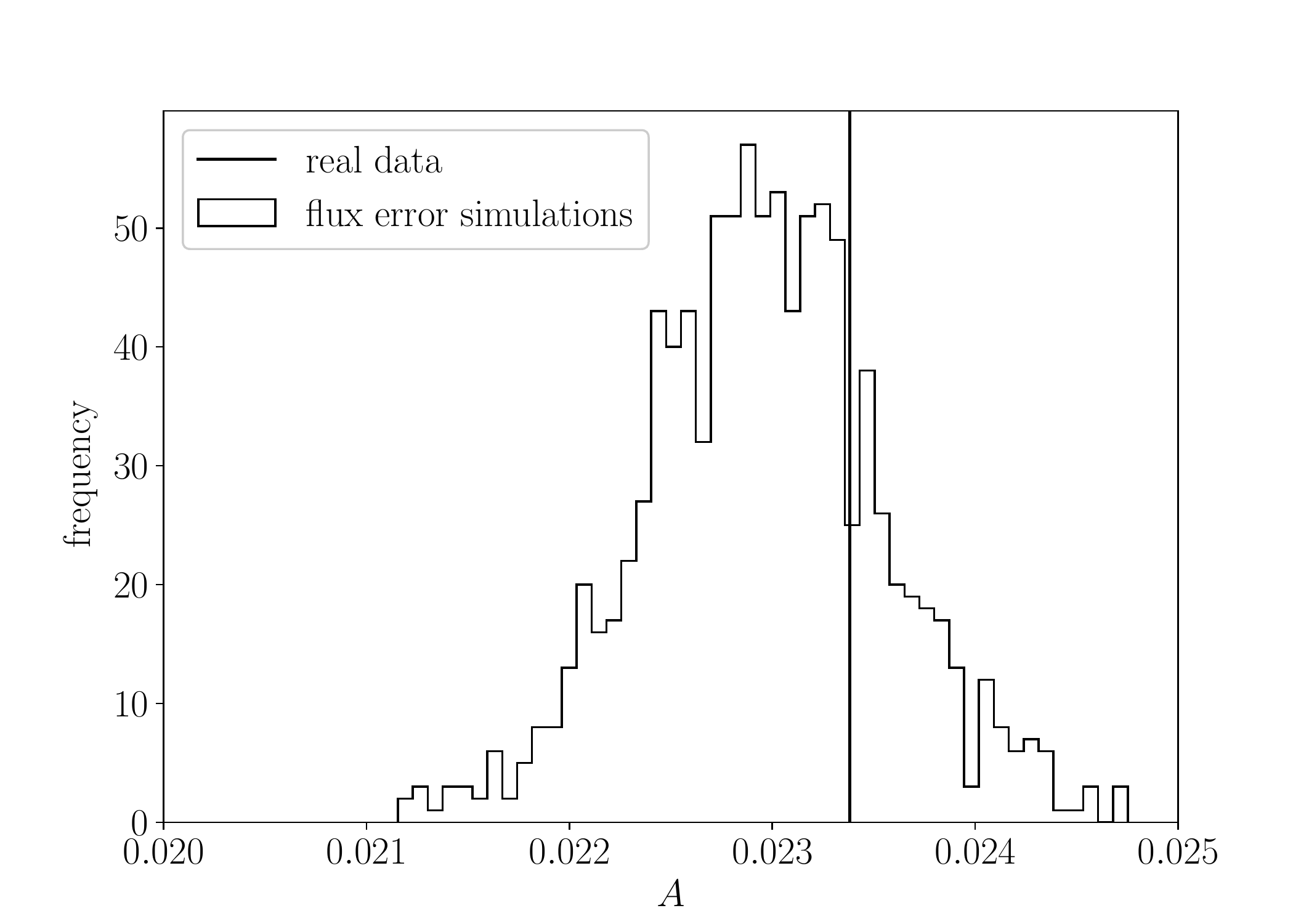}
\includegraphics[width = 8.0cm, height = 5.5cm]{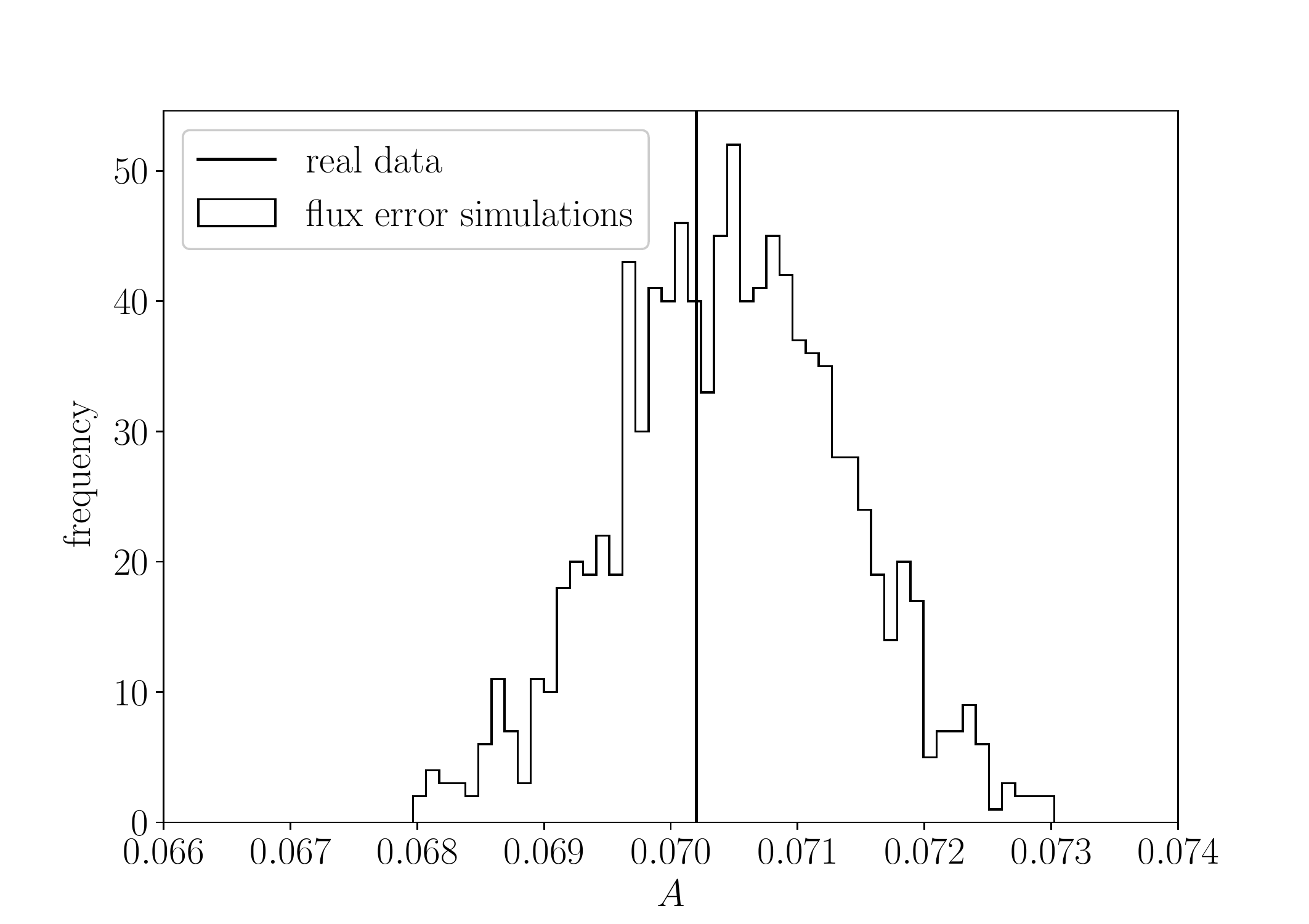}
\caption{Dipole amplitude histograms for the NVSS (left) and TGSS (right) realisations produced following \eqref{eq:sigmaS}. Vertical lines are the dipole value obtained from the actual data.} 
\label{fig:hist_fluxvar}
\end{figure*}

\begin{table}[!t]
\begin{tabular}{ccccc}
\hline
\hline &&&&\\
Survey & Flux range (mJy) & ${A_{\rm obs}}$ & $(l,b)$ & ref.\\ &&&&\\
\hline
\hline
TGSS & $100<S<5000$ & $0.070 \pm 0.004$ & $(243.00^{\circ} \pm 12.00^{\circ}, 45.00^{\circ} \pm 3.00^{\circ})$ & This work \\
NVSS & $20<S<1000$ & $0.023 \pm 0.004$ & $(253.12^{\circ} \pm 11.00^{\circ}, 27.28^{\circ} \pm 3.00^{\circ})$ & This work \\
\hline
\hline
&$S>20$      & $0.021 \pm 0.006$ & $(244.69^{\circ} \pm 27.00^{\circ}, 41.18^{\circ} \pm 29.00^{\circ})$ & \cite{blake02} \\ 
&$20<S<1000$ & $0.021 \pm 0.005$ & $(252.22^{\circ} \pm 10.00^{\circ}, 42.74^{\circ} \pm 9.00^{\circ})$ & \cite{singal11} \\
NVSS & $S>15$  & $0.027 \pm 0.005$ & $(213.99^{\circ} \pm 20.00^{\circ}, 15.30^{\circ} \pm 14.00^{\circ})$ & \cite{gibelyou12} \\
(other work) &$S>25$      & $0.019 \pm 0.005$ & $(248.47^{\circ} \pm 19.00^{\circ}, 45.56^{\circ} \pm 9.00^{\circ})$ & \cite{rubart13} \\
&$S>20$      & $0.010 \pm 0.005$ & $(256.49^{\circ} \pm 9.00^{\circ}, 36.25^{\circ} \pm 11.00^{\circ})$ & \cite{tiwari14} \\
&$S>20$      & $0.012 \pm 0.005$ & $(253.00^{\circ}, 32.00^{\circ})$ & \cite{tiwari16a} \\
&$S>10$      & $0.019 \pm 0.002$ & $(253.00^{\circ} \pm 2.00^{\circ}, 28.71^{\circ} \pm 12.00^{\circ})$ & \cite{colin17} \\ 
\hline
\hline
\end{tabular}
\caption{Measured dipole amplitude ${A_{\rm obs}}$ and direction $(l,b)$ (galactic coordinates) of the TGSS and NVSS catalogues, as obtained from our delta-map \eqref{eq:delta}. Previous results for  NVSS are also cited for comparison.} 
\label{tab:dipole_realdata} 
\end{table}


\begin{figure*}[!t]
\centering
\includegraphics[scale=0.52]{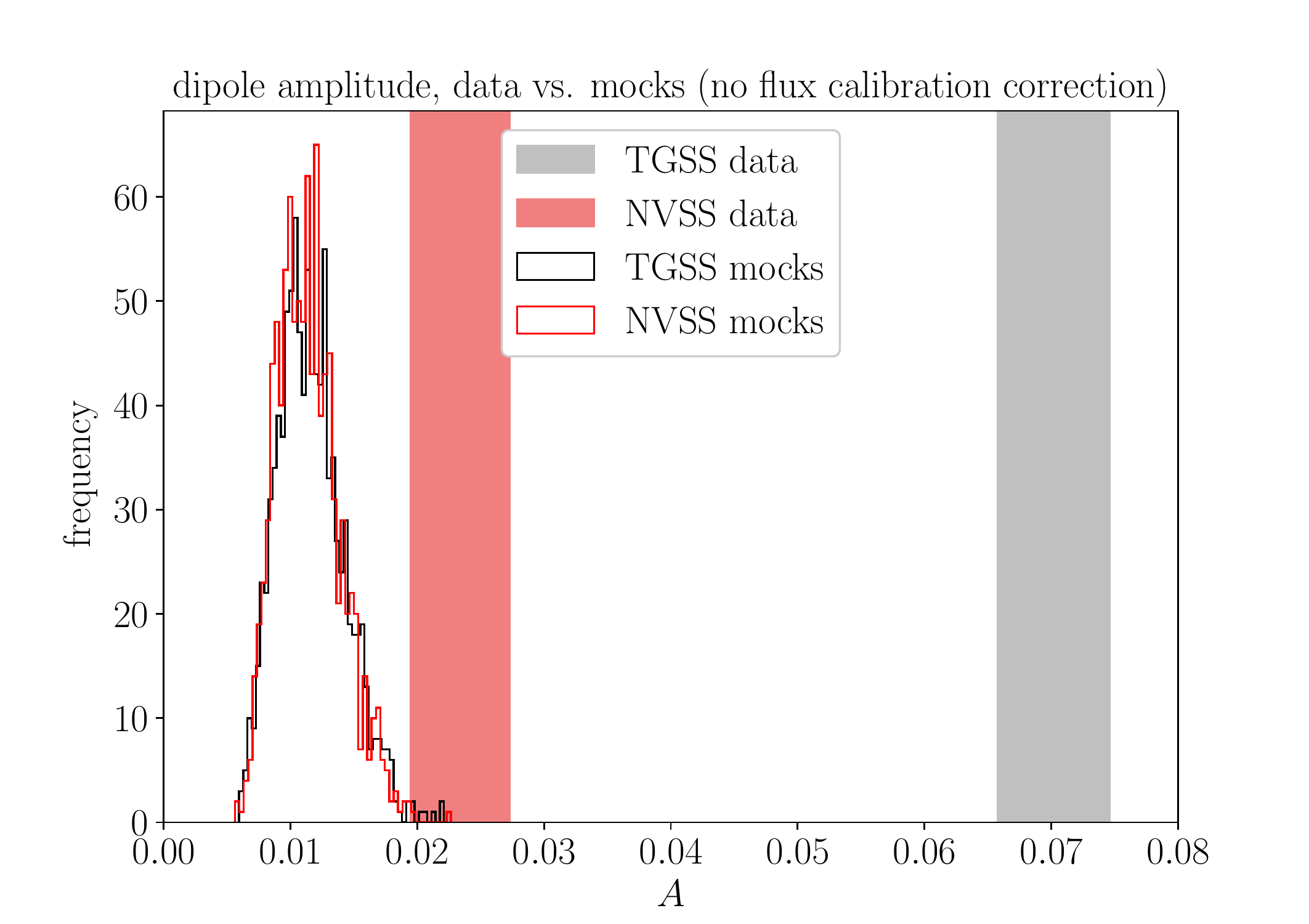}\\
\includegraphics[scale=0.52]{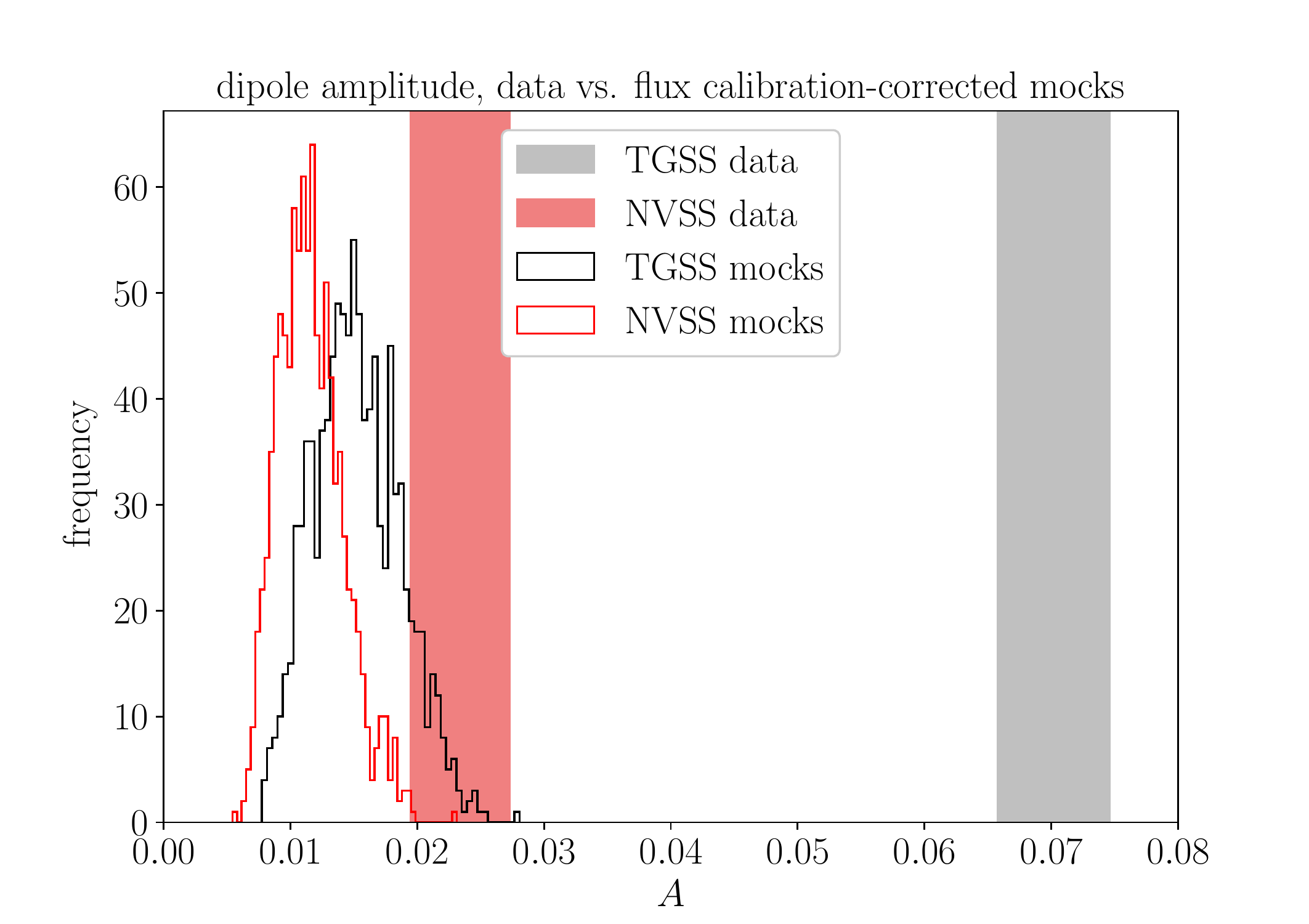}
\caption{Comparison between the dipole amplitude from NVSS (red) and TGSS (black) real and mock catalogues, as obtained from the delta-map estimator -- for the uncorrected (top) and corrected (bottom) flux calibration. Vertical shaded bars indicate the observed values within the error bars, as given by \eqref{aobs}. Histograms show the results from 1,000 realisations.} 
\label{fig:dipole_data_vs_mocks}
\end{figure*}

Figure~\ref{fig:deltamap_realdata} and Table~\ref{tab:dipole_realdata} show the delta-map results for the TGSS and NVSS dipoles. The histograms in Fig.~\ref{fig:hist_fluxvar} provide the dipole amplitude distribution when performing the realisations with flux errors as in \eqref{eq:sigmaS}. The observed amplitudes for each catalogue, along with the uncertainties given by twice the standard deviation obtained from Fig.~\ref{fig:hist_fluxvar}, are
\ba \label{aobs}
A_{\rm obs}^{\rm NVSS} = {0.0234}\pm 0.0039\,, \qquad
A_{\rm obs}^{\rm TGSS} = {0.0702}\pm 0.0044\,.
\ea
The measured dipole directions are
\ba \label{dobs}
(l,b)_{\rm obs}^{\rm NVSS} &=& (253.12^{\circ} \pm 11.00^{\circ}, 27.28^{\circ} \pm 3.00^{\circ})\,,\nonumber \\
(l,b)_{\rm obs}^{\rm TGSS}& =& (243.00^{\circ} \pm 12.00^{\circ}, 45.00^{\circ} \pm 3.00^{\circ})\,.
\ea
The uncertainties correspond to the angular distance at which $\Delta(\theta)$ given in \eqref{eq:delta} falls below 95\% of its maximum.
The TGSS dipole direction is  consistent with that of NVSS, and with the CMB [as given in \eqref{eq:dip_amp_nvss}].

Table~\ref{tab:dipole_realdata} also cites previous NVSS results for comparison. 
Our NVSS results are consistent with previous estimates for the amplitude and its uncertainties -- in particular with~\cite{singal11}, which adopted the same flux range, $20<S<1000 \, \mathrm{mJy}$ (although the mask was different).  Our analysis thus confirms the tension between the amplitudes of the CMB dipole and the radio count dipole, that has been previously reported for the NVSS. 
The dipole directions are all broadly consistent. The uncertainties in direction show a wide variation, with ours similar to the lower end of estimates, such as~\cite{singal11}.

The TGSS dipole should agree, within statistical errors, with that of NVSS. This is the case for the direction -- but the TGSS dipole amplitude is much larger and quite strongly inconsistent with that of the NVSS.

\section{Measured dipoles: mock catalogues}

Figure~\ref{fig:dipole_data_vs_mocks} compares the delta-map dipoles measured from the actual data with the dipoles extracted from the mock catalogues. For the mocks, we compare the results without and with the flux calibration corrections \eqref{eq:corr_N}--\eqref{fct}. We find that
\ba\label{eq:A_mock_CL1}
\mbox{no flux calibration corrections:}\quad && A_{\rm mock}^{\rm NVSS} = {0.0112}_{~-0.0035}^{~+0.0048}\,, \quad
A_{\rm mock}^{\rm TGSS} = {0.0114}_{~-0.0037}^{~+0.0050} \,, \\ 
&& \nonumber \\
\label{eq:A_mock_CL2}
\mbox{with flux calibration corrections:} \quad && A_{\rm mock}^{\rm NVSS} = {0.0114}_{~-0.0036}^{~+0.0050}, \quad
A_{\rm mock}^{\rm TGSS} = {0.0150}_{~-0.0050}^{~+0.0062} \,.
\ea
The central value corresponds to the median dipole amplitude, and the error bars are at 95\% CL. Comparing with \eqref{aobs}, we see that the observed NVSS dipole is $\sim\!2$ times bigger than the predicted median amplitude, while for TGSS it is almost $\sim\!5$ times bigger.

How many of the 1,000 realisations predict a dipole amplitude greater than the observed amplitude? Even after including the contributions from galaxy clustering, Poisson noise, and flux calibration errors, only 2 NVSS realisations satisfy the condition $A_{\rm mock} > A_{\rm obs}$ given the shaded line shown in Fig.~\ref{fig:dipole_data_vs_mocks}, which takes into account the source flux errors. These results imply that there is a tension between the expected and the measured dipole anisotropy in the NVSS radio counts. For the TGSS survey, the even higher  dipole amplitude suggests the presence of persisting systematic errors in the data.

\section{Discussion and conclusions}

\begin{figure*}[!t]
\includegraphics[width = 5.5cm, height = 8.0cm, angle=90]{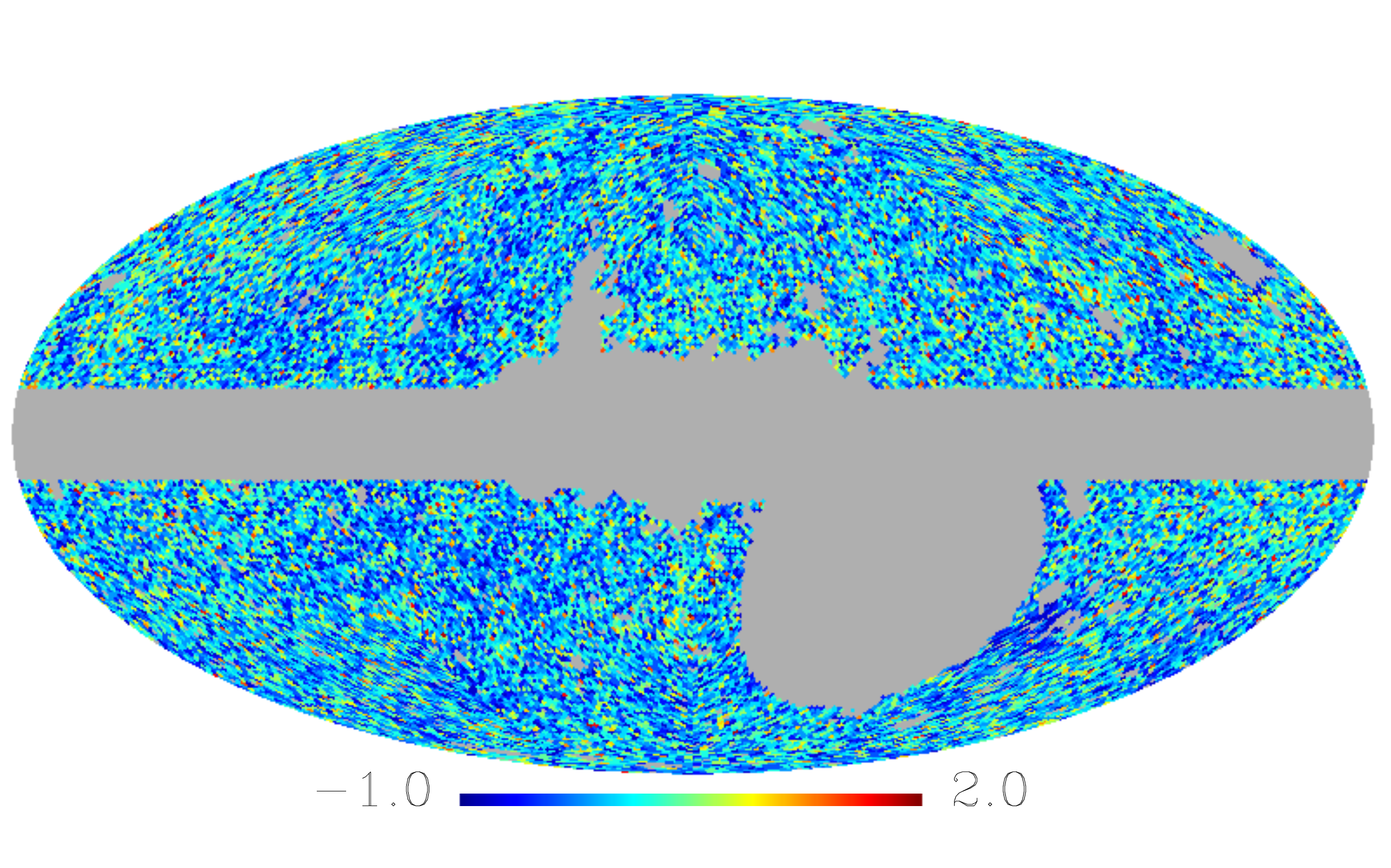}
\includegraphics[width = 5.5cm, height = 8.0cm, angle=90]{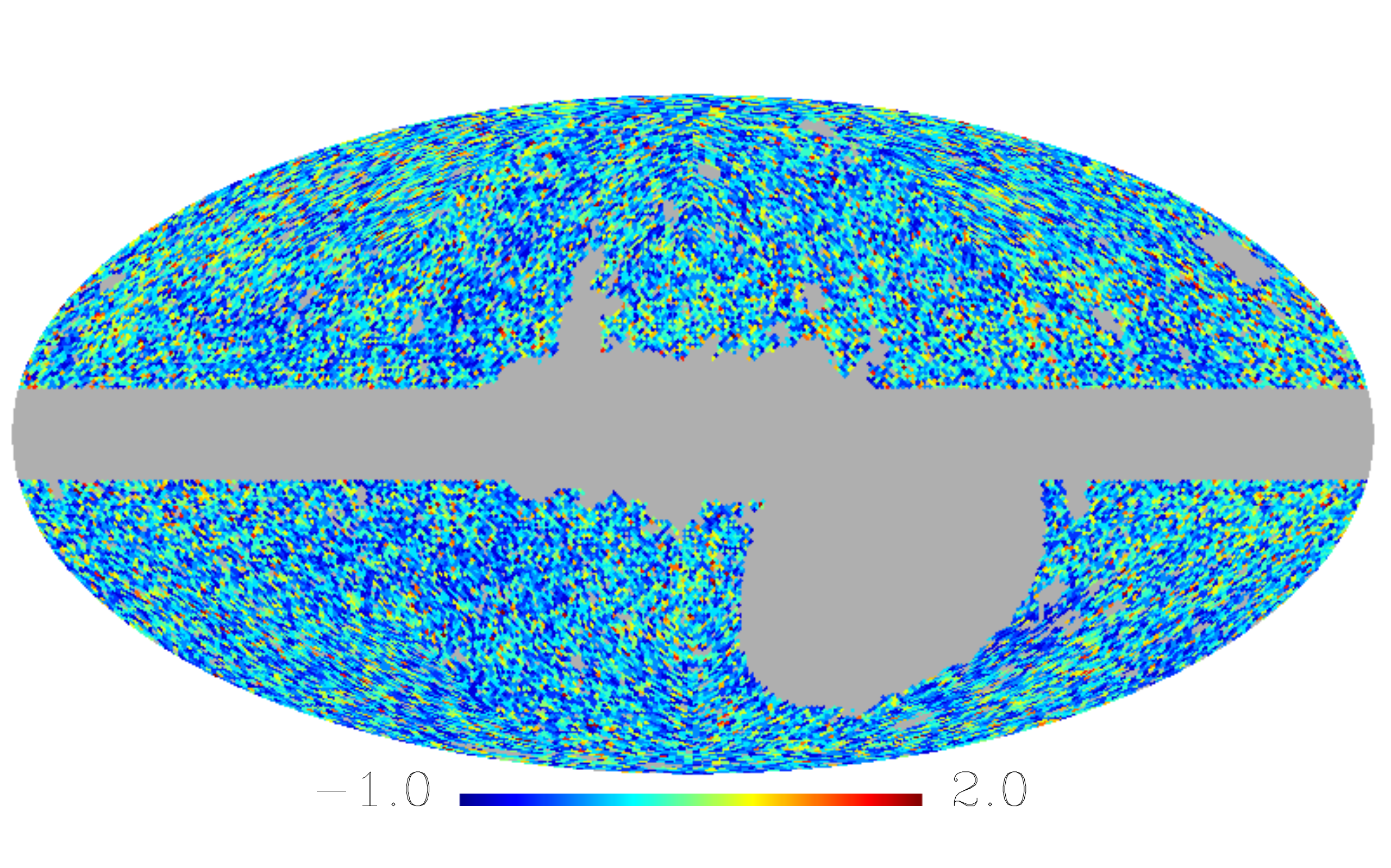}\\
\includegraphics[width = 5.5cm, height = 8.0cm, angle=90]{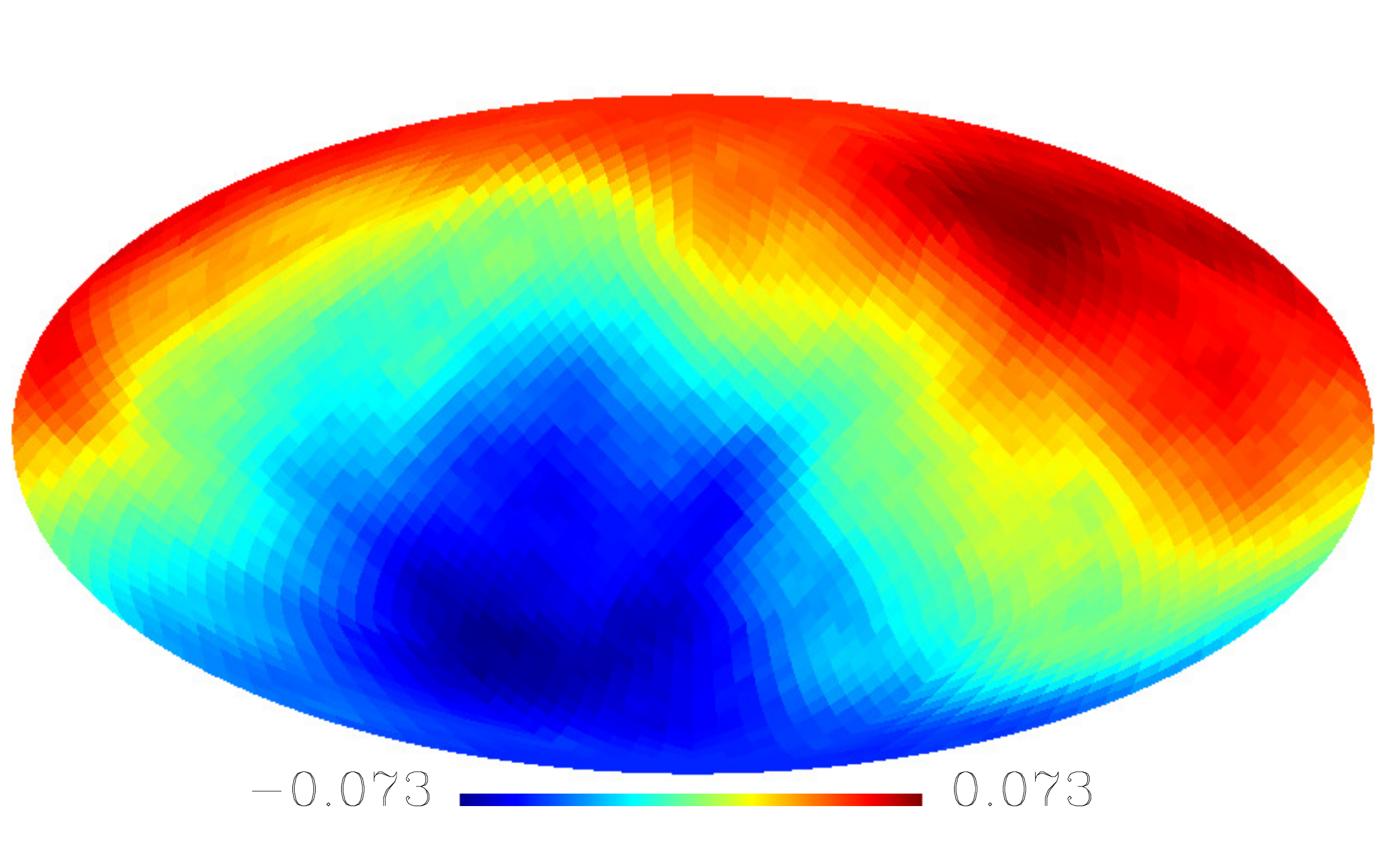}
\includegraphics[width = 5.5cm, height = 8.0cm, angle=90]{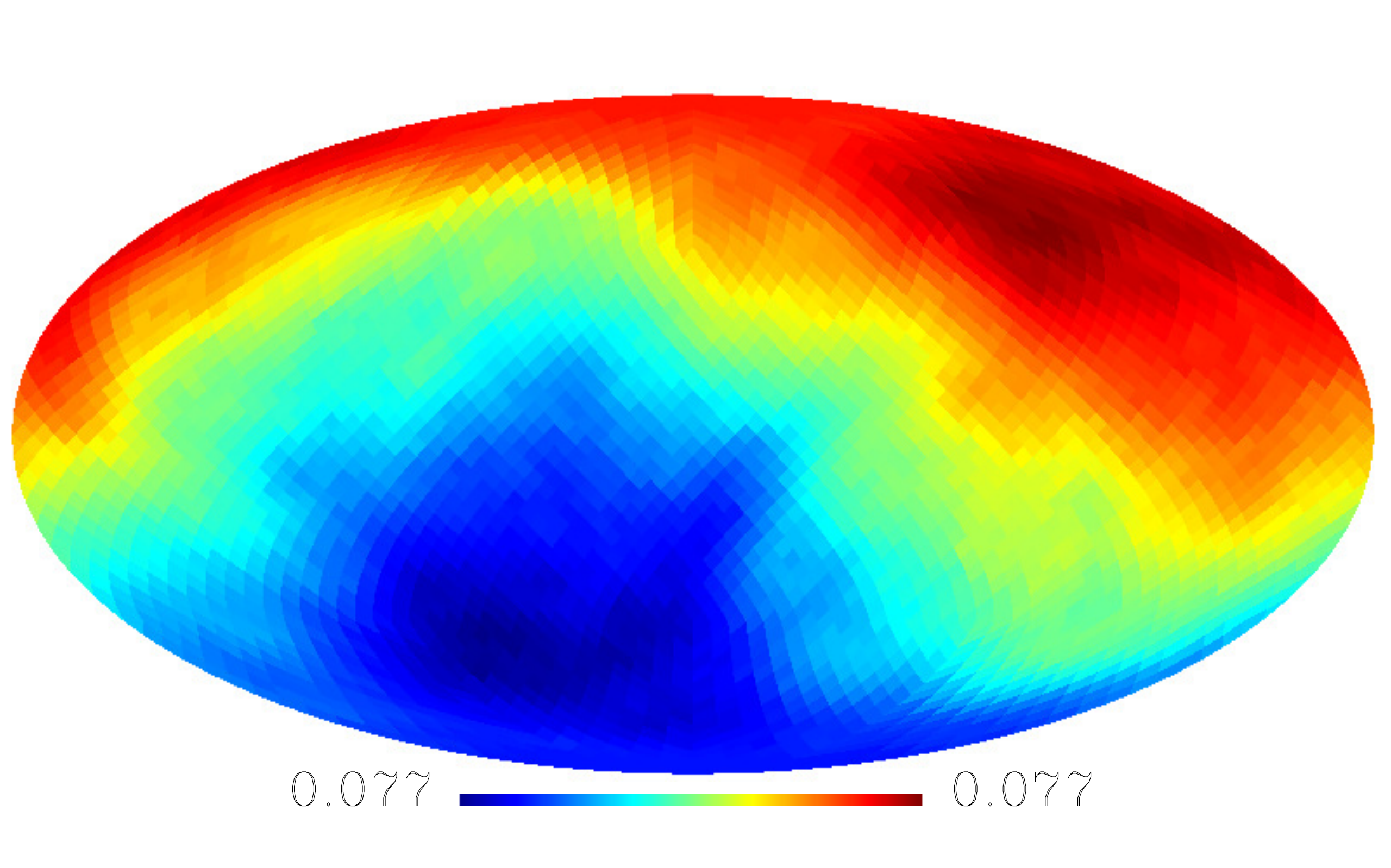}
\caption{Upper: Pixelised TGSS number count map at $150<S<5000$ mJy (left) and $200<S<5000$ mJy (right). Lower: Corresponding delta-map results.} 
\label{fig:nummap}
\end{figure*}

As we can see in Fig.~\ref{fig:deltamap_realdata} and Table~\ref{tab:dipole_realdata}, the dipole directions obtained in the two catalogues are indeed very close to the CMB kinematic dipole, i.e., $(l,b) = (264^{\circ},48^{\circ})$. This is a striking result,  providing another confirmation of the agreement between the CMB and the radio continuum dipole directions, but now with agreement at low and high radio frequencies. By contrast, the dipole amplitudes are inconsistent with each other due to an anomalously large dipole in the TGSS, which is in strong tension with mock realisations.
  
Other analyses reported a tension between the measured and predicted NVSS dipole amplitudes, and our results are consistent with most of them. It is not possible to explain this tension purely in terms of well-known local contaminants (e.g. low-$z$ bright radio sources, nearby clusters, galactic contamination, regions exhibiting high rms flux noise etc.), since most of them have been taken into account, as in~\cite{colin17}. Among  previous analyses, the work of~\cite{tiwari16a} showed that the tension between the estimated and expected NVSS dipole could be somewhat alleviated by taking large-scale galaxy clustering into account. We followed a different method to include the effects of clustering: we used the {\sc flask} code to produce sets of 1,000 lognormal NVSS and TGSS mocks, based on the $\Lambda$CDM angular power spectrum, and on the redshift distribution obtained from the S$^3$ simulations. We adopted a galaxy bias that scales with redshift, the same sky coverage as the actual data-sets -- in addition, we took into account direction-dependent flux calibration errors.For the data, we enforced  flux ranges that ensure completeness of TGSS and remove very bright sources in NVSS, adopted a strict masking scheme,  and incorporated uncertainties in source fluxes. Nevertheless, as shown in Fig.~\ref{fig:dipole_data_vs_mocks}, the mock data is not able to reproduce the observed dipole amplitude $A_{\rm obs}$ within $99.5$\% CL for both surveys, even when the flux  errors are accounted for.

Since different approaches have been used to probe the NVSS dipole, and all of them produce quite similar results, this tension is unlikely to arise from the methods of analysis. However, the tension in the NVSS case is not strong enough to support a claim of violation of the Cosmological Principle. It is more likely to be due to insufficient number density in the NVSS, as shown in Table~\ref{tab:SNR}, and partly because of the large-scale structure and declination-related flux calibration errors. 

For the TGSS dipole amplitude, the tension between measurement and predictions is much stronger. Furthermore, it is inconsistent with the NVSS amplitude, even though the dipole directions are consistent. This inconsistency is already apparent in the angular power spectra of Fig.~\ref{fig:Cls}, which show the excess power in TGSS  on scales $\ell\lesssim 30$.
There are indications that the TGSS data has position-dependent errors in flux density on few degree scales~\cite{hurley17}, and we used information about the TGSS pointings\footnote{H. Intema, private communication.} to model such errors in the mock data. These errors do affect the dipole estimate, as can be seen in the broadening of the TGSS distribution in Fig.~\ref{fig:dipole_data_vs_mocks} (bottom). However, the effect is much smaller than what is required to explain the dipole discrepancy. 

For the measurements, we also included uncertainties in source fluxes, but the effects were small and did not change the results noticeably. Finally, we investigated the effect of the flux cut in TGSS. When we raised the flux cut to $S>150$ and $S>200\,$mJy, the measured dipole amplitude was even {\em larger}, while the direction changed negligibly (see Fig.~\ref{fig:nummap}).

We conclude from these results that (a) there are remaining systematics in the TGSS data, but (b) these systematics appear not to be cured by raising the flux cut and incorporating errors in flux calibration and source fluxes. In addition, it is striking that the direction of the TGSS dipole is consistent with that of NVSS and the CMB. 
 
Further work is needed to analyse systematics in the current data-sets. This will also be valuable for future radio surveys, which will have the number density to make a high signal-to-noise measurement of the dipole direction and amplitude (see, e.g.,~\citep{crawford09, ska15, maartens17}), allowing us to test the Cosmological Principle with much higher confidence. 

\newpage
\noindent{\bf Acknowledgements:}\\
We thank Song Chen, Subir Sarkar and Dominik Schwarz for useful comments. Comments from an anonymous editor and referee helped us to improve the paper. We are grateful to Huib Intema for sharing additional information on the TGSS data.
The authors were supported by the South African SKA Project and the National Research Foundation of South Africa (Grant Nos. 75415 and 84156). RM was also supported by the UK STFC, Grant ST/N000668/1.
Some of our results made use of the HEALpix package~\cite{gorski05}. 


\end{document}